\documentclass[11pt,a4paper]{article}

\usepackage{amsmath}
\usepackage{graphicx}
\usepackage{epsfig}
\usepackage{amsfonts}
\usepackage{amssymb}
\usepackage{placeins}
\usepackage{cases}
\usepackage[titletoc,toc,title]{appendix}
\usepackage{epstopdf}

\usepackage{accents}

\newcommand{\X}{\mathbf{X}}
\newcommand{\Z}{\mathbf{Z}}
\newcommand{\F}{\mathbf{F}}

\newcommand{\ve}{\varepsilon}

\newcommand{\bg}{\mathbf{\gamma}}

\newcommand{\yp}[1]{{#1}}
\newcommand{\dw}[1]{{#1}}

\newcommand*\diff{\mathop{}\!\mathrm{d}}

\begin{document}

\title{Recent Advances in Coupled Oscillator Theory}

\author{
Bard Ermentrout$^{1}$, Youngmin Park$^{2}$ and Dan Wilson$^{3}$}

\maketitle

\setlength{\parindent}{0pt}
$^{1}$Dept of Mathematics, University of Pittsburgh, Pittsburgh, PA 15260 USA \\
$^{2}$Department of Mathematics Goldsmith 218 Mailstop 050 Brandeis University, 415 South Street Waltham, MA 02453, USA\\
$^{3}$Dept of Electrical Engineering and Computer Science, University of Tennessee, Knoxville, TN 37996, USA

\begin{abstract}
We review the theory of weakly coupled oscillators for smooth systems. We then examine situations where application of the standard theory falls short and illustrate how it can be extended. Specific examples are given to non-smooth systems with applications to the Izhikevich neuron.  We then introduce the idea of isostable reduction to explore behaviors that the weak coupling paradigm cannot explain.  In an additional example, we show how bifurcations that change the stability of phase locked solutions in a pair of identical coupled neurons can be understood using the notion of isostable reduction.
\end{abstract}

\tiny
Subjects: Mathematical Neuroscience, Applied Mathematics, Mathematical Biology

Keywords: Weakly coupled oscillators, Isostable coordinates, Phase reduction

\normalsize
\setlength{\parindent}{15pt}
\section{Introduction to Weak Coupling}

Because of its generality and its wide applications in physics, chemistry, and biology, the theory of weakly coupled oscillators has been widely developed and applied over the last forty years.  Starting with Kuramoto (reprinted in \cite{kuramoto03}) and Neu \cite{neu79}, there has been a great deal of development of the theory both in terms of mathematical rigor (\cite{ermentrout1984frequency,wcnn}) and in applications \cite{physd,acebron}. In this paper, we will briefly review the original theory of weakly coupled oscillators and then describe several different extensions that (1) allow for systems with discontinuities and (2) extend beyond weak coupling to include slowly decaying amplitude terms. 


\subsection{General Theory for Weak Coupling}
We begin with just a pair of weakly coupled oscillators in order to show the procedure and how to get the associated coupling functions. Generalizations to $N$ oscillators will be shown afterward. We also describe approaches to delay equations and to PDE models that use the same ideas. We consider a pair of oscillators:

\begin{equation}
\label{eq:wc2}
\frac{dX_j}{dt} = F(X_j) + \epsilon G_j(X_j,X_k), \quad j=1,2;k=3-j,
\end{equation}    
where $0<\epsilon\ll 1$ is a small parameter. $F:R^m\to R^m$ will be assumed to be sufficiently smooth. We assume that the ODE, $X'=F(X)$ has an asymptotically stable $T-$periodic function, $U(t)$. That is, consider the linearized equation:
\[
L(t)v := \frac{dv}{dt}-A(t)v=0,
\]
where $A(t)=D_XF(X)_{X=U(t)}$. Solutions to this equation have the form, $v(t)=e^{\lambda_k t} P_k(t)$ where $P_k(t+T)=P_k(t)$. We assume that there is a simple $\lambda_1=0$ eigenvalue with $P_1(t)=U'(t)$ and that all the remaining $\lambda_k,k=2,\ldots,m$ have negative real parts. The quantities, $\nu_k=\exp(\lambda_k T)$ are called the Floquet multipliers. Thus,  $|\nu_k|<1$ for $k>1.$  There are two approaches to studying the dynamics of the coupled system. The geometric approach makes successive changes of variables and then applies the theory of averaging \cite{neu79,plateaus}.  The more straightforward, though less intuitively appealing, method uses an analytic approach and direct perturbation; often called the adjoint method \cite{wcnn}.  While this method was rigorously formulated by Malkin \cite{wcnn}, it has been widely used as a formal perturbation method.  We will start with the space of square-integrable periodic functions with the inner product
\[
\langle u(t),v(t)\rangle = \int_0^T u(t)\cdot v(t)\ \diff t.
\]   
With this inner product, the linear operator, $L$ has an adjoint operator
\begin{equation}
\label{eq:z}
L^*(t) v = -\frac{dv}{dt} - A(t)^Tv,
\end{equation}
where $A^T$ is the transpose of $A$.  Since $L$ has a one-dimensional nullspace, spanned by $U'(t)$, the adjoint operator also has a one-dimensional nullspace spanned by $Z(t)$, $L^*(t) Z(t)=0$ with the normalization, $Z(t)\cdot U'(t)=1.$   There are several different ways to solve for $Z(t)$, numerically and analytically. One can solve the adjoint equation by solving the appropriate boundary value problem, or by integrating it backwards in time with some random initial data.  A more recent method of forward integration uses the Koopman operator and Fourier averages and extends the notion of isochrons beyond simple periodic orbits \cite{mauroy}. For smooth systems away from bifurcations, backward integration works fine; near bifurcations and for stiff systems, it is better to use the boundary value method or forward integration.   Finally, it is assumed that the Fredholm alternative holds. That is $L(t)u=b(t)$ has a bounded periodic solution if and only if $\langle Z(t),b(t)\rangle=0.$  With these preliminaries in mind, we now derive the coupling functions for a pair of weakly coupled limit cycle oscillators.  We introduce a fast time, $s=t$ and a slow time scale, $\tau=\epsilon t$ and look for solutions to Eq. (\ref{eq:wc2}) that have the form:
\[
X_j(t) = X_{0,j}(s,\tau) + \epsilon X_{1,j}(s,\tau) + \ldots ,
\]
where at each step, $X_{l,j}(s+T,\tau)=X_{l,s}(s,\tau)$. The order 1 equation is
\[
\frac{\partial X_{0,j}}{\partial s} = F(X_{0,j}),
\]
which has a solution $X_{0,j}(s,\tau)=U(s+\theta_j(\tau))$ where $\theta_j(\tau)$ is the (at this point) arbitrary phase shift.  The order 1 equation is
\[
L(s+\theta_j) X_{1,j} = -U'(s+\theta_j)\frac{\partial \theta_j}{\partial\tau} + G_j(U(s+\theta_j),U(s+\theta_k)).
\]
Applying the solvability condition, we obtain:
\begin{equation}
\label{eq:phase2}
\frac{\partial \theta_j}{\partial \tau} = H_j(\theta_k-\theta_j),
\end{equation}
where
\[
H_j(\phi) = \frac{1}{T} \int_0^TZ(t)\cdot G_j(U(s),U(s+\phi))\ \diff s
\]
defines the {\em coupling functions.}  Before analyzing the coupled system, we say a few words about the coupling functions.  Clearly $H_j(\phi+T)=H_j(\phi)$, so that they are $T-$periodic.  Their form depends on both the nature of the coupling, $G_j$ and also on the shape of $Z$. The theory of weak coupling has frequently been applied to networks of spiking neurons (see later in this paper) in which case the coupling has one of two forms:
\begin{equation}
\label{eq:syn}
G_j(X_j,X_k) = g_{jk}s_k(t) (E - V_j(t)),
\end{equation}
synaptic coupling, where $g_{jk} s_k(t)$ is the synaptic conductance and $V_j(t)$ is the post-synaptic potential of the neuron. Note that when synaptic coupling is used, the two oscillators interact through perturbations to their voltage components. Let $Z_v(t)$ be the voltage component of the adjoint solution. In a neural context, $Z_v(t)$ is called the {\em infinitesimal phase resetting curve} (iPRC) and can be measured in real neurons by applying timed current pulses.  For synaptic coupling in this form, 
\[
H_{jk}(\phi) = \frac{g_{jk}}{T}\int_0^T Z_v(t)s(t+\phi) (E-V(t))\ \diff t,
\]   
where $V(t)$ ($s(t)$) is the voltage (synaptic) component of the oscillator, $U(t)$.  
 The other form of coupling is called linear diffusive:
\begin{equation}
\label{eq:diff}
G_j(X_j,X_k) = D_{jk} (X_k-X_j).
\end{equation}
Synaptic coupling is linear in $s$ so that if we know it for, say $s=\sigma(t)$, then if we apply a linear filter, to $\sigma(t)$, it commutes with the coupling function. Specifically let
\[
H_\sigma(\phi) = \frac{1}{T} \int_0^T \sigma(t+\phi) Z_v(t)(E-V(t))\ \diff t,
\]
and let $s(t)=\int_0^\infty f(t')\sigma(t-t')\ \diff t'$ be a filtered version of $\sigma$. Then
\[
H_s(\phi) = \int_0^\infty f(t') H_\sigma(\phi-t')\ \diff t'.
\]  
In particular, if the filter is just a delay, $f(t)=\delta(t-t_d)$, then the coupling function is just, $H(\phi-t_d)$, a phase-shift. These considerations demonstrate how important the timing of interactions are with respect to the shape of the coupling function.  Many authors have studied how different model parameters affect the shape of the coupling function, and notably, the shape of the iPRC \cite{netoff}. For example \cite{vvr} showed that there were transitions between synchrony and other phase locked patterns as the time scale of the synapses changed (specifically, they studied the shapes of the coupling functions for $f(t)=\alpha^2 t e^{-\alpha t}$ as $\alpha$ varied.    

For diffusive coupling
\[
H_{jk}(\phi)= \frac{1}{T} \int_0^T Z(s)\cdot D_{jk}(U(s+\phi)-U(s))\ \diff s.
\]
Note that for diffusive coupling, $H_{jk}(0)=0$. If the diffusion is scalar, that is $D_{jk}=d_{jk}I$, then, $H_{jk}(\phi)=d_{jk}(h(\phi)-h(0))$ with
\[
h(\phi)=\frac{1}{T}\int_0^T Z(s) U(s+\phi)\ \diff s.
\]   
 
Given that we have the coupling functions, we can ask what they tell us about the behavior of the pair of oscillators.  Let $\phi=\theta_2-\theta_1$. Then
\[
\frac{d\phi}{d\tau} = H_2(-\phi)-H_1(\phi) :=C(\phi).
\]
Zeros of $C(\phi)$, $\phi_0$  correspond to phaselocked solutions to the coupled system and if $C'(\phi_0)<0$ (resp $>0$), the locked solution is stable (unstable).  If the oscillators are identical, i.e. $H_1=H_2=H$, then $C(\phi)$ is proportional to the odd part of $H(\phi)$ and so there are always the roots, $\phi=0$, synchrony, and $\phi=T/2$, anti-phase (There can, of course, be other roots as well, but they will always occur symmetrically in pairs due to the fact that $C(\phi)$ is an odd function).  

\begin{figure}
\includegraphics[width=5in]{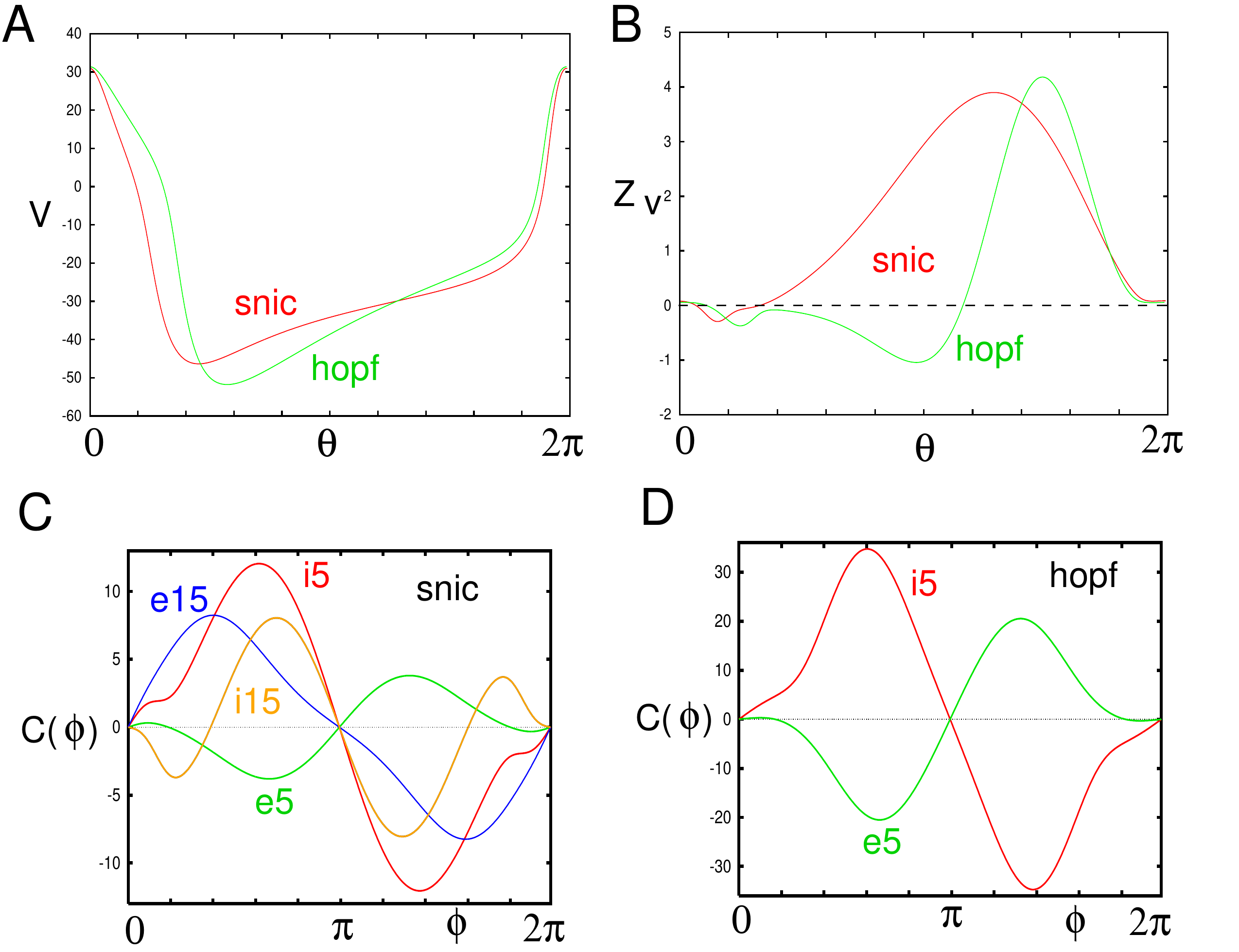}
\caption{The Morris Lecar Model. (A) The voltages for parameters near a SNIC or Hopf bifurcation. (B) The solutions to the adjoint equation in each case. (C) The interaction functions $C(\phi)$ in different scenarios for the SNIC. E (I) excitatory (inhibitory) and $\tau=5,15$ msec; so E15 is excitatory with $\tau=15$, e.g.  (D) Same as (C) for the Hopf scenario; $\tau=15$ differs only in the amplitudes and is not shown.}
\label{fig:mlcf}
\end{figure}

\subsection{Example} As an example of what determines the shape of the coupling functions, we consider the behavior of the Morris-Lecar model in two different scenarios. We choose parameters (as in Table 3.1 in \cite{erme10}) to be near the saddle-node infinite cycle (SNIC) bifurcation with applied current $I_{app}=45$ or near the sub-critical Hopf bifurcation with $I_{app}=92$. The currents are chosen so that the oscillators in each case have a frequency of about 10 Hz.  Coupling is via synapses as in Eq. (\ref{eq:syn}).  We choose excitatory ($E=0 mV$) or  inhibitory ($E=-75 mV$) coupling. The synaptic variables satisfy:
\[
\frac{ds}{dt}=-s/\tau + 1/(1+\exp(-(V-20)/3)),
\]
and we choose $\tau=5,15$ msec.  Fig. \ref{fig:mlcf}A shows the action potential ($V$) over one cycle of the oscillation. The zero phase is set to be the peak of $V(t).$  The two different traces correspond to the SNIC and the Hopf bifurcations. Both oscillators have an uncoupled frequency of 10 Hz and  the action potentials are similar in shape.  In panel B, we show the solutions to the adjoint equation, (\ref{eq:z}) for the two different cases. $Z_v$ is positive everywhere except in a small window near $\theta=0$ for the SNIC. This is a general property of systems near a SNIC \cite{ermentrout1996type}.  In contrast, for the Hopf case, $Z_s$ has a large region of phase-delay that is half the period. These differences in shape matter with respect to the coupling function.   Fig. \ref{fig:mlcf}C,D show the coupling functions, $C(\phi)$ for different types of synapses (excitatory and inhibitory) and different decay times (5,15 msec) in the SNIC (left, C) and the Hopf (right, D) parameters.  Intersection with the axis with negative (positive) slopes correspond to stable (unstable) phase relationships between pairs of oscillators.  At 5 msec delay, excitatory (e5,green) synapses lead to a coupling function with a stable phase difference, $\phi \approx \pm \pi/5$  that is neither synchronous ($\phi=0$) or anti-phase, ($\phi=\pi$). However, once the synapses slow down to $\tau=15$ msec, the stable phase-shift is $\phi=\pi$(e15,blue). Similarly, with inhibition at 5 msec decay (i5,red), $\phi=\pi$ is the only attractor. However, at slower values of decay ($\tau=15$, i15, orange), there is bistability between synchrony ($\phi=0$) and anti-phase ($\phi=\pi$). For the Hopf case, excitatory synapses lead to stable synchrony (or near synchrony), (e5, green) while inhibitory synapses lead to anti-phase (i5,red).  These figures show that both the nature of the coupling and the shape of $Z$ play an important role in determining the phase-locking properties of symmetrically coupled oscillators.

The fact that a pair of identical oscillators always leads to an odd effective coupling function (the even terms don't matter) would lead one to believe that we could assume $H$ is an odd function.    However, once more than two oscillators are connected, then the even component of the coupling function does matter, both in the form of the phase-locked solutions and their stability.  Note that for identical oscillators, $C'(\phi_0)=-(H'(-\phi_0)+H'(\phi_0))$ so that if $H'(\pm\phi_0)>0$, then we have stability. (In particular, this is clear for $\phi_0=0,T/2.$)  For positive scalar diffusive coupling, synchrony is always stable since $h'(0)=(1/T)\int_0^T Z(s)U'(s)\ ds =1.$  \cite{nakao2016} reviews methods of phase reduction applied to partial differential equations as well as to delay equations.   In all cases, the theory is essentially the same, but it becomes necessary to find the function $Z$ for a PDE or functional equation.  There are several technical difficulties that center around finding the correct adjoint equations and dealing with certain boundary terms in the inner product.  We will see this type of issue raise its head in the section on nonsmooth oscillators.   

\subsection{Networks.}   
It is now clear how to generalize the pair of weakly coupled oscillators to networks.  We will assume that coupling in the network has the form, $G_j(X_1,\ldots,X_N)=W_j(X_j)+\sum_{k=1}^N g_{jk} G(X_j,X_k)$ for simplicity, where $W_j$ represents some weak heterogeneity. Letting $\omega_j=(1/T)\int_0^T Z(s)W_j(U(s))\ ds$, and $H(\phi)=(1/T)\int_0^T Z(s)G(U(s),U(s+\phi))$ we obtain
\begin{equation}
\label{eq:phsnet} \theta_j' = \omega_j + \sum_{k=1}^N g_{jk} H(\theta_k-\theta_j),
\end{equation}  
where we have used $\theta'$ to denote the derivative of $\theta$ with respect to $\tau$.  When $g_{jk}=K/N$ and $H(\phi)=\sin\phi$, we recover the classic Kuramoto model. We define a phase-locked solution to Eq. (\ref{eq:phsnet}) as $\theta_j = \Omega \tau + \phi_j$, with $\phi_1=0$ and the others constant.  Ermentrout \cite{stablearray} proved that such a phase-locked solution is stable if $g_{jk}H'(\phi_k-\phi_j)\ge0$ for all $j,k$.  Note that this is a sufficient, but not necessary condition.  

We consider a simple ring of nearest neighbor coupled oscillators and show how the even terms can play a role in both the existence and stability. Consider equation (\ref{eq:phsnet}) with $\omega_j=1$, and $g_{jk}=1$ for $k=j-1,j+1$ and 0 otherwise (with $N+1$ identified with $1$ and $0$ identified with $N$). We will assume $T=2\pi$ with no loss in generality.  A phaselocked solution corresponding to a traveling wave, is $\phi_j=2\pi (j-1)m/N$ along with
\[
\Omega=\Omega_m:=1 + H(2\pi m/N)+H(-2\pi n/M).
\]  
Note that if $H$ is an odd periodic function then $\Omega_m=1$ is independent of $m$.  The relationship between $m$ and $\Omega_m$ is called the dispersion relationship and shows how the network frequency depends on the ``wave'' number, $m/N$.  The linear stability of the wave is easy to determine since the resulting matrix is circulant. Thus, we find that the real part of the eigenvalues are
\[
\mu_{lm} = [H'(2\pi m/N)+H'(-2\pi m/N)](\cos(2\pi l/N)-1).
\]
In other words, for stability, one needs $H'(2\pi m/N)+H'(-2\pi m/N))>0.$

\section{Adjoint Method (Nonsmooth Systems)}

The calculation of the adjoint method for nonsmooth systems is not as straightforward as integrating Equation (\ref{eq:z}). While nonsmooth systems may admit stable limit cycle solutions, potential discontinuities at switching boundaries render Equation \eqref{eq:z} ineffective. Several researchers manage to work around this limitation, determining quantities such as synchronization and phase locking in networks of piecewise linear oscillators.

In 2001, Coombes used a chemically coupled network of piecewise linear planar relaxation oscillators and explored the synchronization properties of the network as a function of fast and slow inhibitory and excitatory synapses \cite{coombes2001phase}. An explicit analysis was possible due to a separation of timescales, weak coupling, and the theory of averaging. \yp{In addition, the piecewise-linear approach of analyzing nodes and networks allows for exact results without the need for reductions that follow from weak interactions \cite{harris2015bifurcations}}.  Coombes et al. (2012) explored the synchronization properties of a linearly coupled network of planar piecewise linear integrate-and-fire (IF) neurons \cite{coombes2012nonsmooth}. To determine stability about the synchronized network state, they introduced a standard perturbation about the synchronous solution and followed the resulting dynamics. They were able to compute the adjoint and iPRC of the planar IF model due to its explicitly solvable nature. Coombes et al. extended the master stability function to the case of coupled piecewise linear oscillators in 2016 \cite{coombes2016synchrony}, \yp{a result which as been further extended to include integrate-and-fire models with state and time-dependent interactions \cite{nicks2018clusters}}.


If the vector field is continuous, the adjoint method can be used directly because the iPRC is continuous across switching boundaries. This property is exploited by Coombes (2008) in calculating the iPRC for gap-junction coupled piecewise linear planar neural models \cite{coombes2008neuronal}. The continuity of the iPRC for continuous $n$-dimensional vector fields is proven in Park et al. (2018) \cite{park2018infinitesimal}, provided that the limit cycle solution transversely cross switching boundaries with nonzero velocity.

\yp{The saltation matrix is a powerful method for analyzing discontinous dynamical systems. Existence and mechanisms of chaos as a result of a discontinuous voltage reset has been shown in a Fitzhugh-Nagumo model by calculating the Lyapunov exponents: a result enabled by the saltation matrix \cite{nobu18}. Assuming continuous solutions in a discontinuous vector field Park et al. (2018) derived the size of discontinuities in the iPRC from first principles \cite{park2018infinitesimal} and showed that the calculation is closely related to the saltation matrix \cite{bernardo2008}. For hybrid systems with discontinuous solutions, Coombes et al. (2012) derived the iPRC for a piecewise linear IF model \cite{coombes2012nonsmooth} by normalizing on each segment away from discontinuities. This method elegantly reproduces the discontinuous iPRC without the need to directly compute the size of the discontinuities. Despite this success, the authors mention that the notion of isochrons, and therefore coupled oscillator theory, is not directly addressed. This problem was handled in 2017 by Shirasaka et al., who rigorously defined isochrons for general hybrid systems, and introduced the phase reduction method for weakly perturbed hybrid systems using the saltation matrix \cite{shirasaka2017phase}.}


\subsection{Weakly Coupled Izhikevich Models}
We now turn to an explicit example of an application of the saltation matrix, which we use to generate the corrected iPRC and predict synchrony in weakly pulse-coupled hybrid limit cycle oscillators. Consider the weakly coupled Izhikevich model \cite{izhikevich2003simple},

\begin{equation}\label{eq:izk}
\begin{split}
\dot\X_i(t) = \left(\begin{matrix}
 \dot v_i\\
 \dot u_i
\end{matrix}\right) &= \left(\begin{matrix}
0.04v_i^2 + 5v_i + 140 - u_i + I\\
a(b v_i-u_i)
\end{matrix}\right)+\ve \yp{|\dot v(T^-)|}\left(\begin{matrix}
\delta(v_{3-i}-30)\\
0
\end{matrix}\right),
\end{split}
\end{equation}
where $i=1,2$, and whenever $v_i\geq 30\text{mV}$, $v_i$ and $u_i$ reset as $v_i \rightarrow c$ and $u_i \rightarrow u_i + d$, respectively. \yp{We choose the coupling function to be the Dirac delta function composed with the voltage variable and scaled by the speed of the voltage variable just before resetting ($|\dot v(T^-)|$). Note that integrating the coupling function on the right-hand side of Equation \eqref{eq:izk} yields
\begin{equation*}
|\dot v(T^-)|\int_0^T \delta(v_{3-i}(t)-30) \diff t = 1,
\end{equation*}
by standard rules of delta function composition, and therefore numerically integrated solutions of Eq. \eqref{eq:izk} receiving weak impulses must increment by order $\ve$}. We use parameters for the regular spiking (RS) neuron: $a=0.02$, $b=0.2$, $c=-65$mV, $d=8$, and $I=10$. The choice of $I$ ensures the existence of a limit cycle solution.

\begin{figure}[ht!]
\centering
 \includegraphics[width=\textwidth]{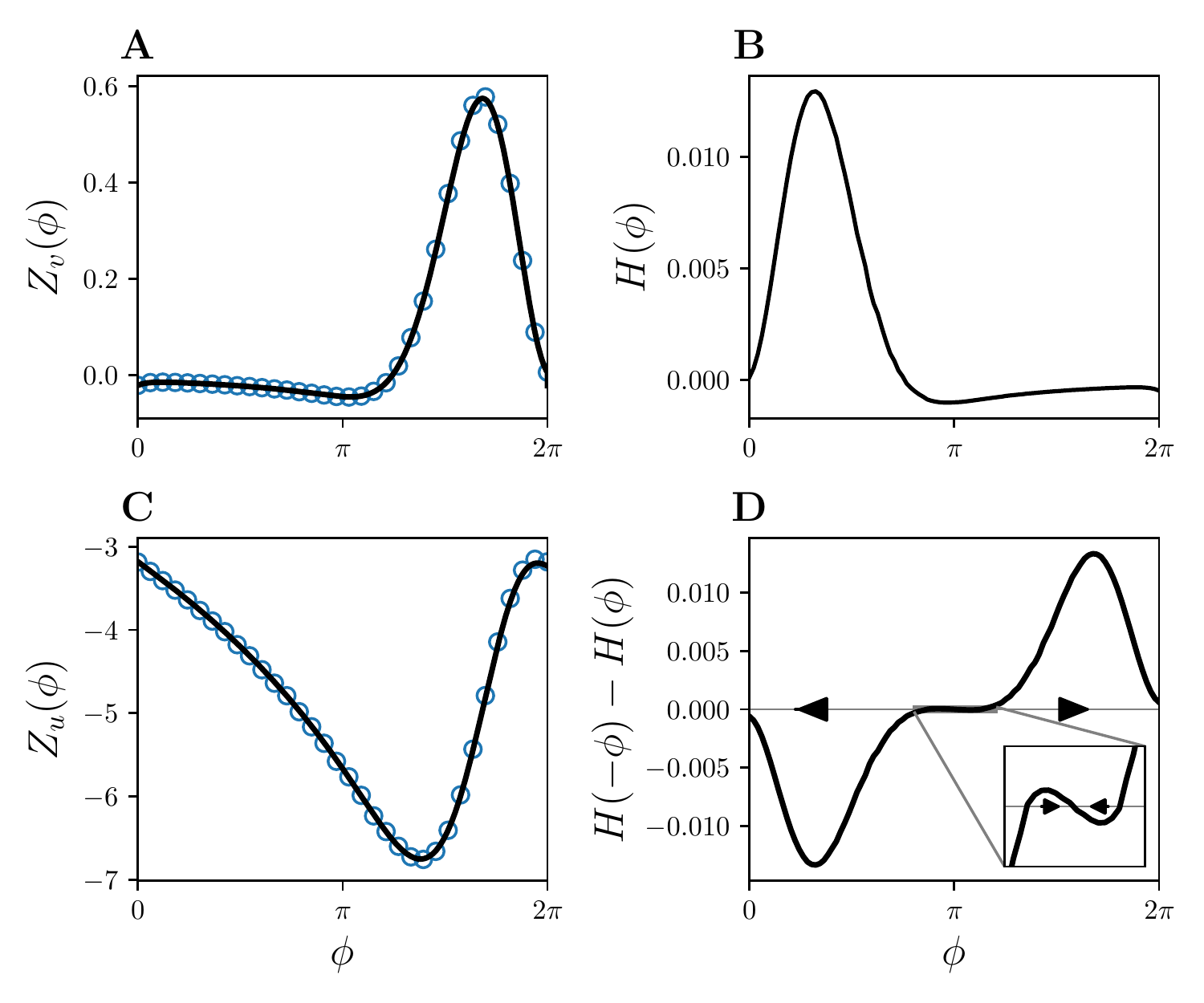}
 \caption{iPRCs and phase functions. A: Adjoint voltage iPRC $Z_v$ (black) and perturbation estimate of the voltage iPRC (blue open circles). B: The interaction function. C: Adjoint $u$ iPRC $Z_u$ (black) and perturbation estimate of the $u$ iPRC (blue open circles). D: Right hand side of the phase dynamics. The synchronous solution is stable, but due to the discontinuity, oscillators synchronize in finite time.}\label{fig:izk_adjoint}
\end{figure}

In order to compute the iPRC, we consider the Izhikevich model in the uncoupled case, where $\ve=0$. This hybrid dynamical system admits a $T$-periodic hybrid limit cycle $\bg(t) = (v^\bg(t),u^\bg(t))^T$. The transition function $\Phi$, which maps solutions $\X(T)=\X^-$ from just before the jump to $\X(T+0)=\X^+$ just after the jump, is given by
\begin{equation*}
 \Phi(\X) = (c, u+d)^T.
\end{equation*}
The switching surface only depends on the voltage variable, thus 
\begin{equation*}
 L(\X) = v-30,
\end{equation*}
where the switching occurs when $v=30$, i.e., when the neuron spikes. The adjoint equation for this problem is given by
\begin{align}\label{eq:adjoint_izk}
 \dot \Z(t) &= -\mathbf{A}^T(t)\Z(t), \quad t \in (0,T),\\
 \Z(t)& = \mathbf{C}^T \Z(t+0), \quad t=T.
\end{align}
These equations are numerically integrated in backwards time and normalized such that $\Z(t) \cdot \F(\bg(t)) = \yp{2\pi/T}$ \cite{shirasaka2017phase} (where $\F(\X)$ is the vector field in Equation \eqref{eq:izk}). The matrix $\mathbf{A} = D\mathbf{F}|_{\bg(t)}$ is the Jacobian matrix evaluated along the limit cycle, and $\mathbf{C}$ is the saltation matrix \cite{leine2013dynamics,bernardo2008,shirasaka2017phase},

\begin{equation}\label{eq:izk_salt}
 \mathbf{C} = D\Phi(\X^-) - [D\Phi(\X^-)\F^- - \F^+]\otimes \left(\frac{\nabla L(\X^-)}{\nabla L(\X^-) \cdot F^-}\right),
\end{equation}
where $\X^-=(30,u^-)^T$ is the solution just before the jump, and $\F^-$ is the vector field just before the jump,
\begin{equation*}
 \F^- = \left(\begin{matrix}
          0.04 30^2 + 5(30) + 140 - u^- + I\\
          a(30b-u^-)
         \end{matrix}\right).
\end{equation*}
The vector $\F^+$ is the vector field just after the jump,
\begin{equation*}
 \F^+ = \left(\begin{matrix}
          0.04 c^2 + 5c + 140 - (u^-+d) + I\\
          a(bc-(u^-+d))
         \end{matrix}\right).
\end{equation*}
For convenience, we write $\F^- = (F_1^-,F_2^-)^T$ and $\F^+ = (F_1^+,F_2^+)^T$. Next, $D\Phi(\yp{\X}^-)$ is the Jacobian matrix of the transition function $\Phi$ evaluated along the limit cycle solution just before the jump,
\begin{equation*}
 D\Phi(\yp{\X}^-) = \left(\begin{matrix}
                     0 & 0\\
                     0 & 1
                    \end{matrix}\right),
\end{equation*}
and finally $\nabla L(\X^-)$ is the gradient of the switching boundary evaluated along the limit cycle just before the jump,
\begin{equation*}
 \nabla L(\X^-) = (1,0).
\end{equation*}
Plugging in these values into Equation \eqref{eq:izk_salt} yields the saltation matrix
\begin{equation*}
 \mathbf{C} = \left(\begin{matrix}
            \frac{F_1^+}{F_1^-} & 0\\
            \frac{F_2^+-F_2^-}{F_1^-} & 1
           \end{matrix}
 \right).
\end{equation*}
With the saltation matrix known, we integrate Equation \eqref{eq:adjoint_izk} backwards in time and normalize the resulting solution at the end of the simulation. The numerically computed adjoint equation $Z_v$ is shown in Figure \ref{fig:izk_adjoint}A,C (solid black), and is plotted against the direct iPRC estimation (open blue circles). Panel A shows the discontinuous voltage iPRC (the discontinuity occurs at $\phi=0$), while panel C shows the continuous $u$ iPRC, $Z_u$. Panels B and D show the functions involved in the phase estimation of the Izhikevich model, which we now explain in detail.

We provide numerical evidence that the classic weak coupling theory applies in the case of non-smooth systems with discontinuous solutions. We remark that an ad-hoc proof of a similar case is covered in Park et al. (2018) \cite{park2018infinitesimal}, but applies to a system with continuous solutions and discontinuous vector fields.

The classic theory of weakly coupled oscillators states that the phase difference $\phi$ between two identical weakly coupled oscillators is given by
\begin{equation*}
 \frac{\diff \phi}{\diff t} = \ve [H(-\phi) - H(\phi)],
\end{equation*}
where the interaction function $H$ is defined as
\begin{equation}\label{eq:interact}
 H(\phi) = \frac{\yp{|\dot v^\bg(T^-)|}}{T}\int_0^T \Z(t) \cdot (\delta(v^\bg(t+\phi)-30),0)^T \diff t,
\end{equation}
where $\bg(t)$ is the uncoupled ($\ve=0$) $T$-periodic limit cycle solution. Thus, the integral is nonzero only when $v^\bg(t+\phi)=30$, which occurs only when $t = T-\phi$, \yp{and} Equation \eqref{eq:interact} reduces to
\begin{equation}\label{eq:izk_interact}
 H(\phi) = Z_v(-\phi)/T,
\end{equation}
and the weakly coupled phase dynamics are entirely determined by the voltage iPRC:
\begin{equation}\label{eq:izk_diff}
 \frac{\diff \phi}{\diff t} = \ve[Z_v(\phi) - Z_v(-\phi)]/T.
\end{equation}
Equation \eqref{eq:izk_interact} is shown in Figure \ref{fig:izk_adjoint}B, and the right hand size of Equation \eqref{eq:izk_diff} is shown in Figure \ref{fig:izk_adjoint}D. Panel D shows that there exist stable fixed points at synchrony ($\phi = 0$), and antiphase ($\phi=\pi$). The basin of attraction of the stable antiphase solution is very small; most initial conditions result in synchrony. Note the discontinuity at $\phi=0$, which comes from the discontinuous voltage iPRC. Solutions in the basin of attraction for synchrony will become synchronous in finite time.

\begin{figure}
\centering
 \includegraphics[width=\textwidth]{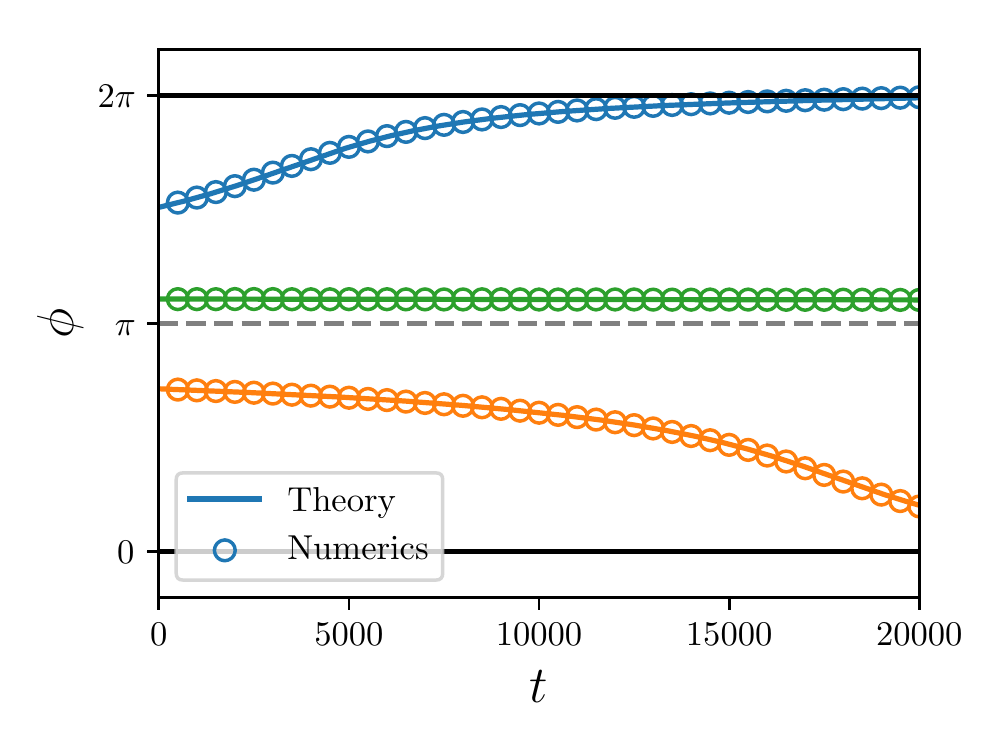}
 \caption{Phase difference dynamics of two weakly pulse-coupled Izhikevich models. Solid curves represent solutions from the theory of weakly coupled oscillators, and open circles represent solutions from the full numerical simulations. Top trace, blue: Initial phase difference at a negative quarter period converges to synchrony. Middle trace, green: Initial phase difference near antiphase stays near antiphase. Bottom trace, orange: Initial phase difference near a quarter period converges to synchrony. $\ve = 0.1$.}\label{fig:izk_timecourse}
\end{figure}

Example solutions leading to these stable fixed points are shown in Figure \ref{fig:izk_timecourse}. We show the results of simulating this phase reduction with various initial phases, and find strong agreement between theory (solid lines) and numerics (open circles). In both the full model and the phase reduction, solutions that tend toward synchrony converge in finite time.

\section{Higher Order Approximations of Coupling Functions}

\dw{The investigation of emergent behaviors in coupled populations of oscillators continues to be an active area of research \cite{ashw16}, \cite{bick16}, \cite{totz18}, \cite{piet19}}.   PRCs and associated phase models (\ref{eq:phase2}) provide first order approximations to phase dynamics resulting from small perturbations; reduction technique allows for complicated models to be analyzed in a more tractible coordinate system.  Such strategies are usually adequate to predict and explain the behavior of coupled oscillators when their underlying limit cycles are strongly stable (i.e,~with Floquet exponents that are negative and large in magnitude) and when the forced behavior is robust to perturbations.  However, in situations where the system is near a bifurcation, higher order approximations for the phase reduced dynamics are necessary to predict and explain the resulting behavior.  

\begin{figure}[htb]
\begin{center}
\includegraphics[width=\textwidth]{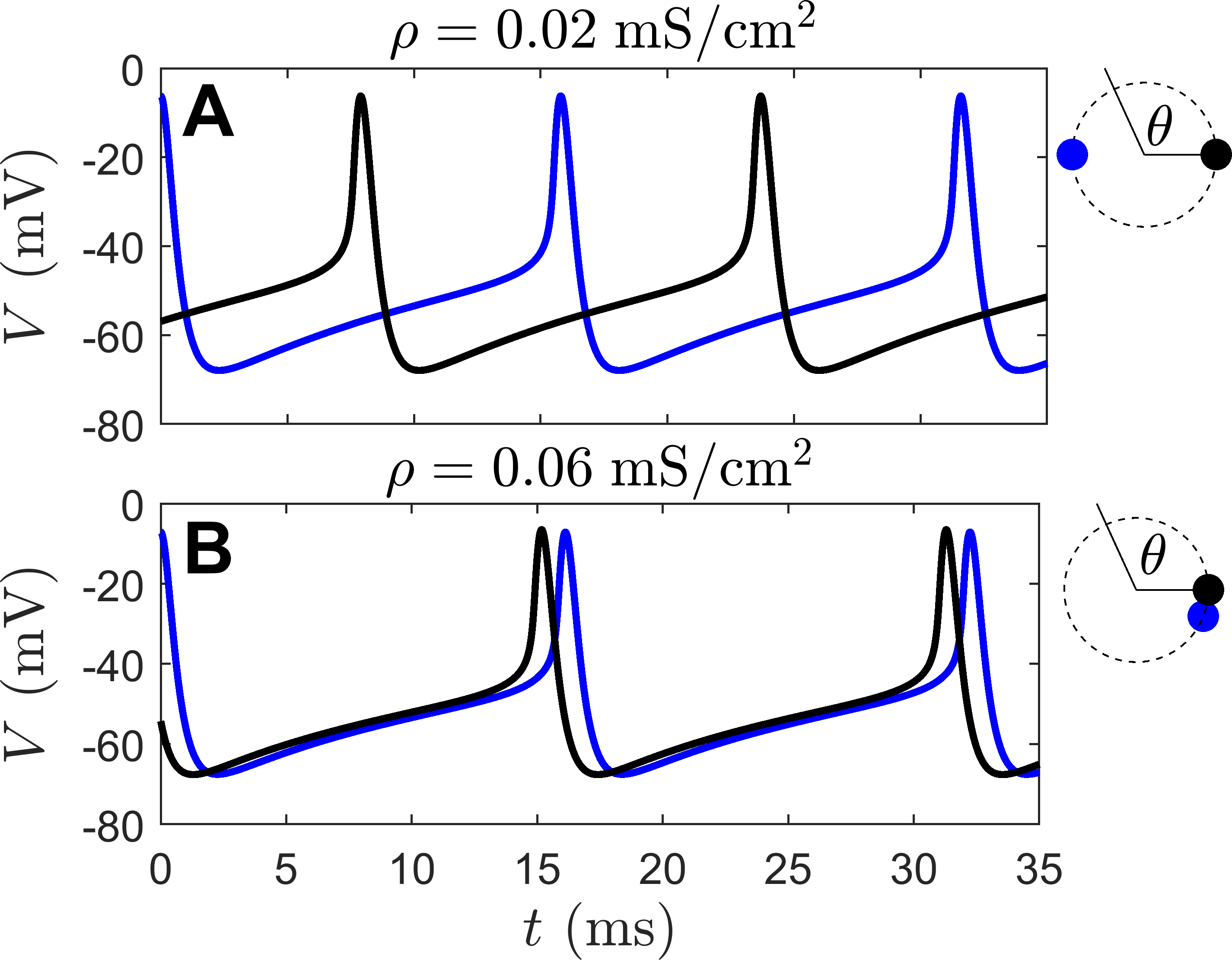}
\end{center}
\caption{Panels A and B show steady state behavior of \eqref{neurmod} for identical initial conditions but with two different coupling strengths.    Note that both neurons are identical.  As shown in the analysis to follow, the steady state in simulations for $\rho = 0.06 \; {\rm mS}/{\rm cm}^2$ results from a saddle node bifurcation.  This behavior can neither be explained nor predicted with standard phase reduction.}
\label{modsim}
\end{figure}

As a concrete example, consider a dynamical model of two identical synaptically coupled thalamic neurons taken from \cite{rubi04}:
\begin{align} \label{neurmod}
C \dot{V}_i &= -I_L(V_i)-I_{Na}(V_i,h_i)-I_K(V_i,h_i)-I_T(V_i,r_i) + I_b - I_{syn}(V_i,s_1,s_2), \nonumber \\
\dot{h}_i &= (h_\infty(V_i)-h_i)/\tau_h(V_i), \nonumber \\
\dot{r}_i &= (r_\infty(V_i) - r_i)/\tau_r(V_i), \nonumber \\
\dot{w}_i&= \alpha(1-w_i)/(1+\exp(-(V_i-V_T)/\sigma_T)) - \beta w_i, \qquad i = 1,2.
\end{align}
Here, $V_i$ is the transmembrane voltage, $h_i$ and $r_i$ are gating variables, $w_i$ determines the synaptic current, $I_L = g_L(V_i - E_L)$, $I_{Na} = g_{Na}m_\infty^3(V_i) h_i (V_i-E_{Na})$, $I_K = g_K(.75(1-h_i))^4(V_i-E_K)$, $I_T = g_T p_\infty^2(V_i) r_i(V_i-E_T)$, $I_b = 3.75 \mu {\rm A}/\mu {\rm F}$ are leak, sodium, potassium, low-threshold calcium, and baseline currents, respectively.   We take conductances $g_L = 0.15$, $g_{Na} = 3$, $g_K = 5$, and $g_T = 10$ ${\rm mS}/{\rm cm}^2$, reversal potentials $E_L = -75$, $E_{Na} = 3$, $E_K = -90$, and $E_T = 0$ mV, and $C = 1 \mu {\rm F}/{\rm cm}^2$.  Synaptic current $I_{\rm syn} = \rho(w_1+w_2)(V_i-V_{syn})$ where $\rho$ determines the magnitude of the coupling, $V_{syn} = -60$ mV, $\alpha = 3 {\rm m s}^{-1}$, $V_T = -20$ mV, $\sigma_T = 0.8$ mV, and $\beta = 0.2{\rm ms}^{-1}$.   All remaining functions are identical to those from \cite{rubi04}.   Simulating \eqref{neurmod} using $\rho = 0.02 \; {\rm mS}/{\rm cm}^2$ and $\rho = 0.06 \; {\rm mS}/{\rm cm}^2$ with initial phases that are nearly identical yields the results shown in Figure \ref{modsim}.  For this model the infinite time behavior depends on the coupling strength itself.  As we will show, this behavior cannot be explained with first order phase reduction techniques alone; higher order corrections must be used.  

Understanding the dynamical behavior in directions transverse to the limit cycle (i.e.,~the amplitude coordinates) is critical to developing higher order approximations of the phase dynamics and there are many possible options for representing both the phase and amplitude coordinates.  For instance, \cite{kuramoto03} and \cite{wils18operat} use hyperplanes to denote surfaces of constant phase as part of a higher order asymptotic expansion, \cite{lets18}  and \cite{wedg13} use a moving orthonormal coordinate frame in the definition of phase-amplitude coordinates, and \cite{cast13}, \cite{wils16isos}, and \cite{shir17} define amplitude coordinates based on Floquet theory.  The coordinates based on Floquet theory have been shown to be particularly useful as they result in relatively simple second order accurate phase-amplitude reduced dynamics \cite{wils17isored} \cite{wils19}.  This strategy will be used in the following analysis to explain the results from Figure \ref{modsim}.  This method of phase-amplitude reduction (sometimes called isostable reduction) is briefly summarized below.

\subsection{Second Order Reduction Using Isostable Coordinates}
The following provides a summary of the work from \cite{wils16isos} and \cite{wils17isored}.  Consider a general equation of the form
\begin{equation} \label{fxeq}
\dot{x} = F(x)  +  g(t),
\end{equation}
where $x\in \mathbb{R}^n$ is the state, the dynamics are given by $F(x)$, and $g(t) \in \mathbb{R}^n = \begin{bmatrix} u(t) & 0 & \dots & 0 \end{bmatrix}^T$  is a small perturbation.  For the moment, we suppose that $u(t) = 0$ and suppose \eqref{fxeq} has a $T$-periodic orbit $x^\gamma(t)$.  Isochrons \cite{guck75} can be used to define phase coordinates $\theta \in [0,2\pi)$ for which $\dot{\theta} = \omega = 2 \pi/T$.   We denote $\Gamma_0$ as the $\theta = 0$ isochron.  By definition, $T$ is the return time from $\Gamma_0$ to $\Gamma_0$.  One can use $\Gamma_0$ as a Poincar\'e surface with associated map
\begin{align}
P:\Gamma_0 \rightarrow \Gamma_0, \nonumber \\
x \mapsto \eta(T,x),
\end{align}
where $\eta$ is the unperturbed flow.  The fixed point of this map, $x_0$, corresponds to the intersection of  $x^\gamma(t)$ and the $\Gamma_0$ surface.  Linearization about this fixed point yields
\begin{equation}
\eta(T,x) = x_0 + J_\eta(x-x_0),
\end{equation}
where $J_\eta$ denotes the Jacobian of $\eta(T,x)$ evaluated at $x_0$.  \dw{ Diagonalization of $J_\eta$ yields eigenvalues $\lambda_k$ with associated left and right eigenvalues $w_k$ and $v_k$, respectively, for $k = 1,\dots,n$.  For every non-unity eigenvalue $\lambda_k$ of $J_\eta$ with equal algebraic and geometric multiplicity, an isostable coordinates $\psi_k$ can be defined as in  \cite{wils16isos}, \cite{wils17isored}}
\begin{equation} \label{isodef}
\psi_k(x) = \lim_{j \rightarrow \infty} \left[ w_k^T  (\eta(t_\Gamma^j,x) - x_0)  \exp (-\kappa_k  t_\Gamma^j) \right],
\end{equation}
where $t_\Gamma^j$ is the $j^{\rm th}$ return time to $\Gamma_0$ under the flow and $\kappa_k = \log(\lambda_k)/T$ is a Floquet exponent.  \dw{In the definition \eqref{isodef}, in the limit as time approaches infinity the decay under the flow $\eta$ matches the growth of $ \exp (-\kappa_k  t_\Gamma^j)$ in the direction specified by $w_k$.  This limiting behavior gives the isostable coordinate $\psi_k(x)$ which is defined for all locations in the basin of attraction of the limit cycle.} Because \eqref{isodef} is defined according to the infinite time convergence of solutions to the limit cycle, one can show (as in \cite{wils16isos}) that under the flow, 
\begin{equation}
\dot{\psi}_k = \kappa_k \psi_k
\end{equation}
for all locations in the basin of attraction of the limit cycle.  In general, $n-1$ isostable coordinates can be defined according to \eqref{isodef} (one for each non-unity eigenvalue of $J_\eta$).  However, for the model considered in the following analysis, all but one Floquet multiplier is very small in magnitude; in this case all other isostable coordinates can be neglected because they decay rapidly and only one isostable coordinate is required \dw{(which will be denoted by $\psi$)} to characterize the behavior transverse to the limit cycle. 

As illustrated in \cite{wils17isored}, starting with a general equation of the form \eqref{fxeq}, one can instead work in phase-isostable reduced coordinates 
\begin{align}\label{order2reduc}
\dot{\theta} &= \omega +  \left[ z(\theta) +   \psi b(\theta)\right]  u(t), \nonumber \\
\dot{\psi} &= \kappa \psi +  \left[ i(\theta) +  \psi  c(\theta) \right] u(t).
\end{align}
Here, $i(\theta)$ is an isostable response curve (analogous to the PRC $z(\theta)$ for the phase variable), and $b(\theta)$ and $c(\theta)$ provide nonlinear corrections to the perturbed dynamics as the system for locations far from the limit cycle.  Methods similar to the adjoint method described in prior sections have been developed for computation of the functions $b(\theta), i(\theta)$, and $c(\theta)$ as detailed in \cite{wils17isored}, and \cite{wils19}.

\subsection{Second Order Accurate Coupling Functions}
Here, we apply the second order isostable reduction methodology to explain the behavior observed in Figure \ref{modsim}.  To begin, we rewrite each neuron from \eqref{neurmod}  in the form \eqref{order2reduc}, where 
\begin{align} \label{thetapsieq}
\dot{\theta}_i &= \omega + \left[  z(\theta_i) + \psi_i b(\theta_i)  \right] u_i(t), \nonumber \\
\dot{\psi}_i &= \kappa \psi_i + \left[ i(\theta_i) + \psi_i c(\theta_i) \right] u_i(t), \quad i = 1,\dots,2,
\end{align}
and
\begin{align} \label{uteq}
u_i(t) &= -I_{\rm syn}(V_i,w_1,w_2) \nonumber \\
&= -\rho(w_1(t)+w_2(t))(V_i(t)-V_{syn}) \nonumber \\
&= -\rho \big( w_1(\theta_1) +  \psi_1 q^w(\theta_1) + w_2(\theta_2) + \psi_2 q^w(\theta_2) \big)\big(  V_i(\theta_i) + \psi_i q^V(\theta_i) - V_{syn}   \big) + \mathcal{O}(\epsilon^3).
\end{align}
\dw{In the above equation, $q(\theta)\in \mathbb{R}^n$ is the eigenfunction associated with the Floquet exponent $\kappa$ for the uncoupled oscillators, and $q^V(\theta) \in \mathbb{R}$ and $q^w(\theta) \in \mathbb{R}$ are defined as individual components of $q(\theta)$ in the coordinates $V$ and $q$, respectively.} As shown in \cite{wils17isored} and \cite{wils19}  as a consequence of Floquet theory one can write $V_i(t) = V(\theta_i (t) ) + \psi_i q^V(\theta_i(t)) + \mathcal{O}(\epsilon^2)$  and $w_i(t) = w(\theta_i(t)) + \psi_i q^w(\theta_i(t)) + \mathcal{O}(\epsilon^2)$.  In the analysis to follow,  $\rho$, $\psi_1$, and $\psi_2$ are assumed to be $\mathcal{O}(\epsilon)$ terms.   Expanding \eqref{thetapsieq} and only retaining terms to leading order $\epsilon^2$ we have,
\begin{align} \label{psitheta2}
\dot{\theta}_1 &= \omega + \rho \left[ h_1(\theta_1,\theta_2) +  \psi_1 h_2(\theta_1,\theta_2) + \psi_2 h_3(\theta_1,\theta_2)\right], \nonumber  \\
\dot{\psi}_1 &= \kappa \psi_1 + \rho \left[ h_4(\theta_1,\theta_2) +  \psi_1 h_5(\theta_1,\theta_2) + \psi_2 h_6(\theta_1,\theta_2)\right], \nonumber  \\
\dot{\theta}_2 &= \omega + \rho \left[ h_1(\theta_2,\theta_1) +  \psi_2 h_2(\theta_2,\theta_1) + \psi_1 h_3(\theta_2,\theta_1)\right],  \nonumber \\
\dot{\psi}_2 &= \kappa \psi_2 + \rho \left[ h_4(\theta_2,\theta_1) +  \psi_2 h_5(\theta_2,\theta_1) + \psi_1 h_6(\theta_2,\theta_1)\right],
\end{align}
where $h_1(\theta_1,\theta_2) = -z(\theta_1)(w(\theta_1)+w(\theta_2))(V(\theta_1) - V_{syn})$, $h_2(\theta_1,\theta_2) = -b(\theta_1)(w(\theta_1)+w(\theta_2))(V(\theta_1)-V_{syn}) - z(\theta_1)q^w(\theta_1)(V(\theta_1)-V_{syn}) - z(\theta_1)(w(\theta_1)+w(\theta_2))q^V(\theta_1)$, $h_3(\theta_1,\theta_2) = -z(\theta_1) q^w(\theta_2)(V(\theta_1)-V_{syn})$, $h_4(\theta_1,\theta_2) = -i(\theta_1)(w(\theta_1)+w(\theta_2))(V(\theta_1) - V_{syn})$, $h_5(\theta_1,\theta_2) = -c(\theta_1)(w(\theta_1)+w(\theta_2))(V(\theta_1)-V_{syn}) - i(\theta_1)q^w(\theta_1)(V(\theta_1)-V_{syn}) - i(\theta_1)(w(\theta_1)+w(\theta_2))q^V(\theta_1)$, and $h_6(\theta_1,\theta_2) = -i(\theta_1) q^w(\theta_2)(V(\theta_1)-V_{syn})$.

By defining new variables $\phi_j = \theta_j - \omega t$ and substituting into \eqref{psitheta2} we have
\begin{align} \label{preavg}
\dot{\phi}_1 &=  \rho \left[ h_1(\phi_1 + t,\phi_2 + t) +  \psi_1 h_2(\phi_1 + t,\phi_2 + t) + \psi_2 h_3(\phi_1 + t,\phi_2 + t)\right], \nonumber  \\
\dot{\psi}_1 &= \kappa \psi_1 + \rho \left[ h_4(\phi_1 + t,\phi_2 + t) +  \psi_1 h_5(\phi_1 + t,\phi_2 + t) + \psi_2 h_6(\phi_1 + t,\phi_2 + t)\right], \nonumber  \\
\dot{\phi}_2 &=  \rho \left[ h_1(\phi_2 + t,\phi_1 + t) +  \psi_2 h_2(\phi_2 + t,\phi_1 + t) + \psi_1 h_3(\phi_2 + t,\phi_1 + t)\right],  \nonumber \\
\dot{\psi}_2 &= \kappa \psi_2 + \rho \left[ h_4(\phi_2 + t,\phi_1 + t) +  \psi_2 h_5(\phi_2 + t,\phi_1 + t) + \psi_1 h_6(\phi_2 + t,\phi_1 + t)\right].
\end{align}
\dw{Recalling that $\psi_1$ and $\psi_2$ are assumed to be $\mathcal{O}(\epsilon)$ terms}, equation \eqref{preavg} can be written in the form $\dot{y} = \epsilon Q(y,t)$.  Additionally  each of the $h_i$ functions from \eqref{preavg} is $T$-periodic so that averaging \cite{sand07}, \cite{guck83} can be applied resulting in 
\begin{align}\label{avgeq}
\dot{\Phi}_1 & = \rho \left[  H_1(\Phi_1-\Phi_2) + \psi_1 H_2(\Phi_1-\Phi_2) + \psi_2 H_3(\Phi_1-\Phi_2) \right], \nonumber \\
\dot{\Psi}_1 & = \kappa \Psi_1 + \rho \left[  H_4(\Phi_1-\Phi_2) + \Psi_1 H_5(\Phi_1 - \Phi_2) + \Psi_2 H_6(\Phi_1-\Phi_2) \right], \nonumber \\
\dot{\Phi}_2 & = \rho \left[  H_1(\Phi_2-\Phi_1) + \Psi_2 H_2(\Phi_2-\Phi_1) + \Psi_1 H_3(\Phi_2-\Phi_1) \right], \nonumber \\
\dot{\Psi}_2 & = \kappa \Psi_2 + \rho \left[  H_4(\Phi_2-\Phi_1) + \Psi_2 H_5(\Phi_2 - \Phi_1) + \Psi_1 H_6(\Phi_2-\Phi_1) \right],
\end{align}
where $H_i(X) = \frac{1}{T}\int_0^T h_i(X+t,t) \diff t$.  Because fixed points of \eqref{avgeq} correspond to periodic solutions of \eqref{preavg} with the same stability \cite{sand07}, \eqref{avgeq} can be used to assess phase locking in \eqref{thetapsieq} (which is in turn used to assess phase locking in \eqref{neurmod}).  Finally, \eqref{avgeq} can be simplified taking $\Upsilon \equiv \Phi_1 - \Phi_2$ to write
\begin{align} \label{finaleq}
\dot{\Upsilon} &= \rho \left[    H_1(\Upsilon)-H_1(-\Upsilon) + \Psi_1(H_2(\Upsilon)-H_3(-\Upsilon)) + \Psi_2(H_3(\Upsilon)-H_2(-\Upsilon))   \right], \nonumber \\
\dot{\Psi}_1 &= \kappa \Psi_1 +  \rho \left[ H_4(\Upsilon) + \Psi_1 H_5(\Upsilon) + \Psi_2 H_6(\Upsilon)    \right], \nonumber \\
\dot{\Psi}_2 &= \kappa \Psi_2 +  \rho \left[ H_4(-\Upsilon) + \Psi_1 H_6(-\Upsilon) + \Psi_2 H_5(-\Upsilon)    \right]. \nonumber \\
\end{align}
We note \eqref{finaleq} is an order $\epsilon^2$ approximation for the phase and isostable dynamics.  If we instead take only an order $\epsilon$ approximation (the usual approach) the resulting phase difference equation would be 
\begin{align}\label{fostab}
\dot{\Upsilon}^{\epsilon} &= \rho\left[  H_1(\Upsilon^{\epsilon})-H_1(-\Upsilon^{\epsilon})  \right], \nonumber \\
\dot{\Psi}_1^{\epsilon} &= \kappa \Psi_1^{\epsilon} + \rho H_4(\Upsilon^{\epsilon}), \nonumber \\
\dot{\Psi}_2^{\epsilon} &= \kappa \Psi_2^{\epsilon} + \rho H_4(-\Upsilon^{\epsilon}),
\end{align}
where $\Upsilon^{\epsilon}$, $\Psi_1^{\epsilon}$, and $\Psi_2^{\epsilon}$ are order $\epsilon$ approximations for the phase difference and isostable coordinates.

\subsection{Results}
\begin{figure}[htb]
\begin{center}
\includegraphics[height=2.1in]{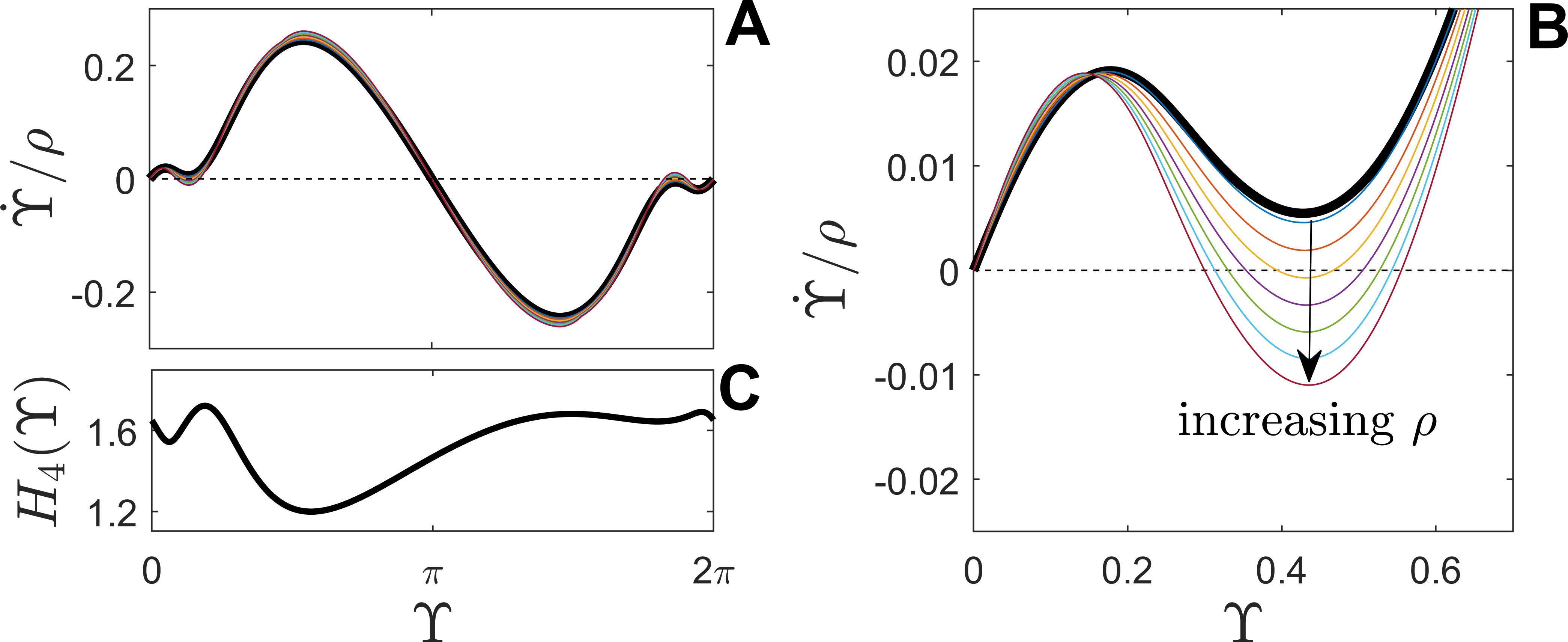}
\end{center}
\caption{ Panels A and B show $\dot{\Upsilon}/\rho$ from \eqref{finaleq} evaluated at the unstable fixed point where $\Upsilon = 0$ and $\Psi_1$ and $\Psi_2$ are determined by $\rho$.  The colored lines show the coupling functions as $\rho$ increases by increments of 0.03.  The black line corresponds to the first order accurate coupling function $H_1(\Upsilon)-H_1(-\Upsilon)$.   While the synchronous solution (resp.,~antiphase) solutions are always stable (resp.,~unstable)  bistability emerges through a saddle node bifurcation as $\rho$ is increased.  This bifurcation cannot be observed in the more well-established $\mathcal{O}(\epsilon)$ accurate reduction strategy  \eqref{fostab}.  Panel C shows $H_4(\Upsilon)$ which determines the order $\epsilon$ accurate behavior of the isostable coordinates in \eqref{finaleq}. }
\label{bifnfig}
\end{figure}

Both neurons from \eqref{neurmod} admit a stable periodic orbit with $T = 15.33$ ms.  For this periodic orbit, the non-unity Floquet multipliers are 0.680, 0.011, and 0.008.  Here, the isostable coordinate $\psi$ corresponds to the direction of slowest decay towards the periodic orbit and the other two directions are neglected because they decay rapidly.  After numerically computing the required functions $z(\theta), i(\theta), b(\theta)$, and $c(\theta)$ as well as terms related to the synaptic coupling from \eqref{uteq}  using methods described in \cite{wils19}, we compute each $h_i$ and subsequent $H_i$ function.   In panels A and B of Figure \ref{bifnfig}, the thick black line shows $H_1(\Upsilon)-H_1(-\Upsilon)$, the first order accurate coupling function from \eqref{fostab}.  Note that the shape of this function has no dependence on the coupling strength, $\rho$, and changing $\rho$ will not alter stable fixed points of \eqref{fostab}.  Panel C shows $H_4(\Psi)$, which influences the order $\epsilon$ dynamics of the isostable coordinates.  Notice that it is strictly positive indicating that increasing $\rho$ will shift the isostable coordinate to more positive values.  The colored lines in panels A and B show $\dot{\Upsilon}/\rho$ from \eqref{finaleq} evaluated at the unstable fixed point $\begin{bmatrix}  \Upsilon & \Psi_1 & \Psi_2 \end{bmatrix} = \begin{bmatrix}  0 & \Psi_1^{fp}(\rho) & \Psi_2^{fp}(\rho) \end{bmatrix}$ as  $\rho$ increases by increments of 0.03.  As $\rho$ increases, both  $\Psi_1^{fp}(\rho)$ and $\Psi_2^{fp}(\rho)$ increase modifying the resulting coupling function.  For all values of $\rho$, there is an unstable fixed point at $\Upsilon = 0$, but as the coupling strength increases, stable (and corresponding unstable) fixed points emerge nearby as the result of a saddle node bifurcation.  For this parameter set, this saddle node bifurcation occurs at $\rho = 0.0481$ at the locations $\begin{bmatrix}  \Upsilon & \Psi_1 & \Psi_2 \end{bmatrix} = \begin{bmatrix}  0.43 &  3.25  &  3.15\end{bmatrix}$ and $\begin{bmatrix}  -0.43 &  3.15  &  3.25\end{bmatrix}$.

The analysis of the reduced equations \eqref{finaleq} agrees well with the observed behavior in the unreduced equations \eqref{neurmod}.  For values of $\rho\leq 0.0463$ the asynchronous state is the only stable configuration.  for $\rho >  0.0463$ an additional stable configuration exists where the neurons fire approximately 1 ms apart which corresponds to a phase difference of about 0.4, similar to the location and coupling strength for which the saddle node bifurcation emerges in the reduced model \eqref{finaleq}.

\section{Conclusions}
Weak coupling theory of oscillators has shown itself to be a powerful tool that can be generally applied to a variety of problems across many areas of science. Here we have shown that extensions to the theory both to nonsmooth systems and \dw{to systems where higher order coupling terms significantly influence the behavior} remain amenable to analysis and allow one to get sharper results when applied to full model equations. While our focus has been primarily on models from theoretical neuroscience, the methods here can be applied to many other fields.

All figure-generation code in this article is available on GitHub at 

https://github.com/youngmp/ermentrout\_park\_wilson\_2019.

\bibliographystyle{plain}




\end{document}